\documentclass[hidelinks]{article}

\usepackage{arxiv}

\usepackage[utf8]{inputenc} 
\usepackage[T1]{fontenc}    
\usepackage{hyperref}       
\usepackage{url}            
\usepackage{booktabs}       
\usepackage{amsfonts}       
\usepackage{nicefrac}       
\usepackage{microtype}      
\usepackage{lipsum}         
\usepackage{graphicx}
\usepackage[square,numbers]{natbib}
\usepackage{doi}
\usepackage{xcolor}
\usepackage{amsmath}
\usepackage{cleveref}       
\usepackage{tikz}
\usepackage{xr}
\usepackage{hyperref}

\title{Time-varying ecological interactions characterise equilibrium and stability}

\newif\ifuniqueAffiliation

\ifuniqueAffiliation 
\author{ 
    \href{https://orcid.org/0000-0001-8930-856X}{\includegraphics[scale=0.06]{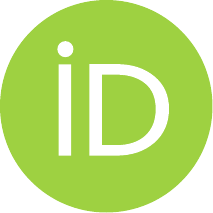}\hspace{1mm}Annalisa Caligiuri} \\
	Institute for Cross-Disciplinary Physics and Complex Systems (IFISC, CSIC-UIB)\\
	  Palma de Mallorca, Spain\\
    \texttt{annalisa.caligiuri@gmail.com}
    \And
    \href{https://orcid.org/0000-0001-5074-1163}
    {\includegraphics[scale=0.06]{orcid.pdf}\hspace{1mm}Emile Emery} \\ SPEC, Université Paris-Saclay, CEA, CNRS
	\\
	  Gif-sur-Yvette, France\\
    \texttt{eemery@protonmail.com}\\
    \And
    \href{https://orcid.org/0000-0001-7871-6066}
    {\includegraphics[scale=0.06]{orcid.pdf}\hspace{1mm}Leonardo Ferreira} \\ Physics Institute, Federal University of Rio Grande do Sul
	\\
	  Porto Alegre, Brazil\\
    \texttt{leop.dsf@gmail.com} \\
    \And
    \href{https://orcid.org/0009-0003-7379-7356}
    {\includegraphics[scale=0.06]{orcid.pdf}\hspace{1mm}Juan García-Castillo} \\
	Institute for Cross-Disciplinary Physics and Complex Systems (IFISC, CSIC-UIB)\\
	  Palma de Mallorca, Spain\\
    \texttt{jantonio@ifisc.uib-csic.es} 
    \And    
	\href{https://orcid.org/0000-0002-1795-6143}{\includegraphics[scale=0.06]{orcid.pdf}\hspace{1mm}Simon D.~Lindner} \\
	Medical University of Vienna\\
	Vienna, Austria \\
	\texttt{lindner.sd@gmail.com}\\
    \And
	\href{https://orcid.org/0009-0006-1789-7741}{\includegraphics[scale=0.06]{orcid.pdf}\hspace{1mm}Javier ~Molina-Hernández} \\
	Universidad Carlos III de Madrid (UC3M),\\
    Grupo Interdisciplinar de Sistemas Complejos (GISC),\\
	Madrid, Spain \\
	\texttt{jamolina@pa.uc3m.es}
    \And
        \href{https://orcid.org/0000-0003-1869-2976}
    {\includegraphics[scale=0.06]{orcid.pdf}\hspace{1mm}Nelson Aloysio ~Reis de Almeida Passos} \\
	{University of Pisa} \\
	{National Research Council} \\
	Pisa, Italy \\
	\texttt{nelson.reis@phd.unipi.it} \\
    \And
    \href{https://orcid.org/0000-0002-2641-2877}{\includegraphics[scale=0.06]{orcid.pdf}\hspace{1mm}Vítor Hugo Ribeiro} \\ State University of Maringá\\
    Maringá, Brazil\\
    \texttt{vitorhibeiro@gmail.com} \\
    \And
    \href{https://orcid.org/0009-0003-2641-886X}{\includegraphics[scale=0.06]{orcid.pdf}\hspace{1mm}Marika Sartore} \\
	Department of Physics, University of Padua,\\
	Padua, Italy \\
	\texttt{marika.sartore@phd.unipd.it}\\
    \And
    \href{https://orcid.org/0000-0001-7276-370X}
    {\includegraphics[scale=0.06]{orcid.pdf}\hspace{1mm}Boxuan Wang} \\
	Sorbonne Universit\'{e}, INSERM, Institut Pierre Louis d'Epid\'{e}miologie et de Sant\'{e} Publique\\
	Paris, France \\
	\texttt{boxuan.wang@iplesp.upmc.fr} \\
    \And
        \href{https://orcid.org/0000-0001-7917-8984}
    {\includegraphics[scale=0.06]{orcid.pdf}\hspace{1mm}Violeta ~Calleja-Solanas} \\
	Doñana Biological Station (CSIC)\\
	Sevilla, Spain \\
	\texttt{violeta.calleja@ebd.csic.es} \\
}
\else
\usepackage{authblk}

\newbox{\orcid}\sbox{\orcid}{\includegraphics[scale=0.06]{orcid.pdf}}

\author[1]{\href{https://orcid.org/0000-0001-8930-856X}{\usebox{\orcid}\hspace{1mm}Annalisa Caligiuri}}
\author[2]{\href{https://orcid.org/0000-0001-5074-1163}{\usebox{\orcid}\hspace{1mm}Emile Emery}}
\author[3]{\href{https://orcid.org/0000-0001-7871-6066}{\usebox{\orcid}\hspace{1mm}Leonardo Ferreira}}
\author[1]{\href{https://orcid.org/0009-0003-7379-7356}{\usebox{\orcid}\hspace{1mm}Juan Garc\'ia-Castillo}}
\author[4]{\href{https://orcid.org/0000-0002-1795-6143}{\usebox{\orcid}\hspace{1mm}Simon D.~Lindner}}
\author[5,11]{\href{https://orcid.org/0009-0006-1789-7741}{\usebox{\orcid}\hspace{1mm}Javier Molina-Hern\'andez}}
\author[6,7]{\href{https://orcid.org/0000-0003-1869-2976}{\usebox{\orcid}\hspace{1mm}Nelson Aloysio Reis de Almeida Passos}}
\author[8]{\href{https://orcid.org/0000-0002-2641-2877}{\usebox{\orcid}\hspace{1mm}V\'itor Hugo Ribeiro}}
\author[9]{\href{https://orcid.org/0009-0003-2641-886X}{\usebox{\orcid}\hspace{1mm}Marika Sartore}}
\author[10]{\href{https://orcid.org/0000-0001-7276-370X}{\usebox{\orcid}\hspace{1mm}Boxuan Wang}}
\author[12]{\href{https://orcid.org/0000-0001-7917-8984}{\usebox{\orcid}\hspace{1mm}Violeta Calleja-Solanas\thanks{\texttt{violeta.calleja@ebd.csic.es}}}}

\affil[1]{Institute for Cross-Disciplinary Physics and Complex Systems (IFISC, CSIC-UIB), Palma de Mallorca, Spain}
\affil[2]{SPEC, Universit\'e Paris-Saclay, CEA, CNRS, Gif-sur-Yvette, France}
\affil[3]{Physics Institute, Federal University of Rio Grande do Sul, Porto Alegre, Brazil}
\affil[4]{Medical University of Vienna, Vienna, Austria}
\affil[5]{Universidad Carlos III de Madrid (UC3M), Madrid, Spain}
\affil[6]{University of Pisa, Pisa, Italy}
\affil[7]{National Research Council, Pisa, Italy}
\affil[8]{State University of Maring\'a, Maring\'a, Brazil}
\affil[9]{Department of Physics, University of Padua, Padua, Italy}
\affil[10]{Sorbonne Universit\'e, INSERM, Institut Pierre Louis d'Epid\'emiologie et de Sant\'e Publique, Paris, France}
\affil[11]{Grupo Interdisciplinar de Sistemas Complejos (GISC), Madrid, Spain}
\affil[12]{Do\~nana Biological Station (CSIC), Sevilla, Spain}

\hypersetup{
pdftitle={Busy Bees: Working Paper},
pdfsubject={q-bio.NC, q-bio.QM},
pdfauthor={Violeta ~Calleja-Solanas},
pdfkeywords={Temporal networks, Ecological networks, Equilibrium, Structural Stability},
}

\begin{document}
\maketitle

\begin{abstract}
	Ecological communities are composed of species interactions that respond to environmental fluctuations. Despite increasing evidence of temporal variation in these interactions, most theoretical frameworks remain rooted in static assumptions. Here, we develop and apply a time-varying network model to five long-term ecological datasets spanning diverse taxa and environments. Using a generalized Lotka-Volterra framework with environmental covariates, we quantify temporal rewiring of interspecific interactions, asymmetry patterns, and structural stability. Our results reveal contrasting dynamics across ecosystems: in datasets with rich temporal resolution, interaction networks exhibit marked rewiring and shifts in cooperation-competition ratios that correlate with environmental stress, consistent—though not always linearly—with the stress-gradient hypothesis. Conversely, in datasets with coarser temporal sampling, networks retain constant interaction sign structure and remain in cooperation-dominated regimes. These findings highlight the importance of temporal resolution and environmental context in shaping ecological coexistence.
\end{abstract}


\section{Introduction}
\begin{figure}[ht]
    \centering
    \begin{minipage}[l]{0.43\linewidth}
        \begin{tikzpicture}
            \node at (0,0) {
            \includegraphics[width=1.0\textwidth]{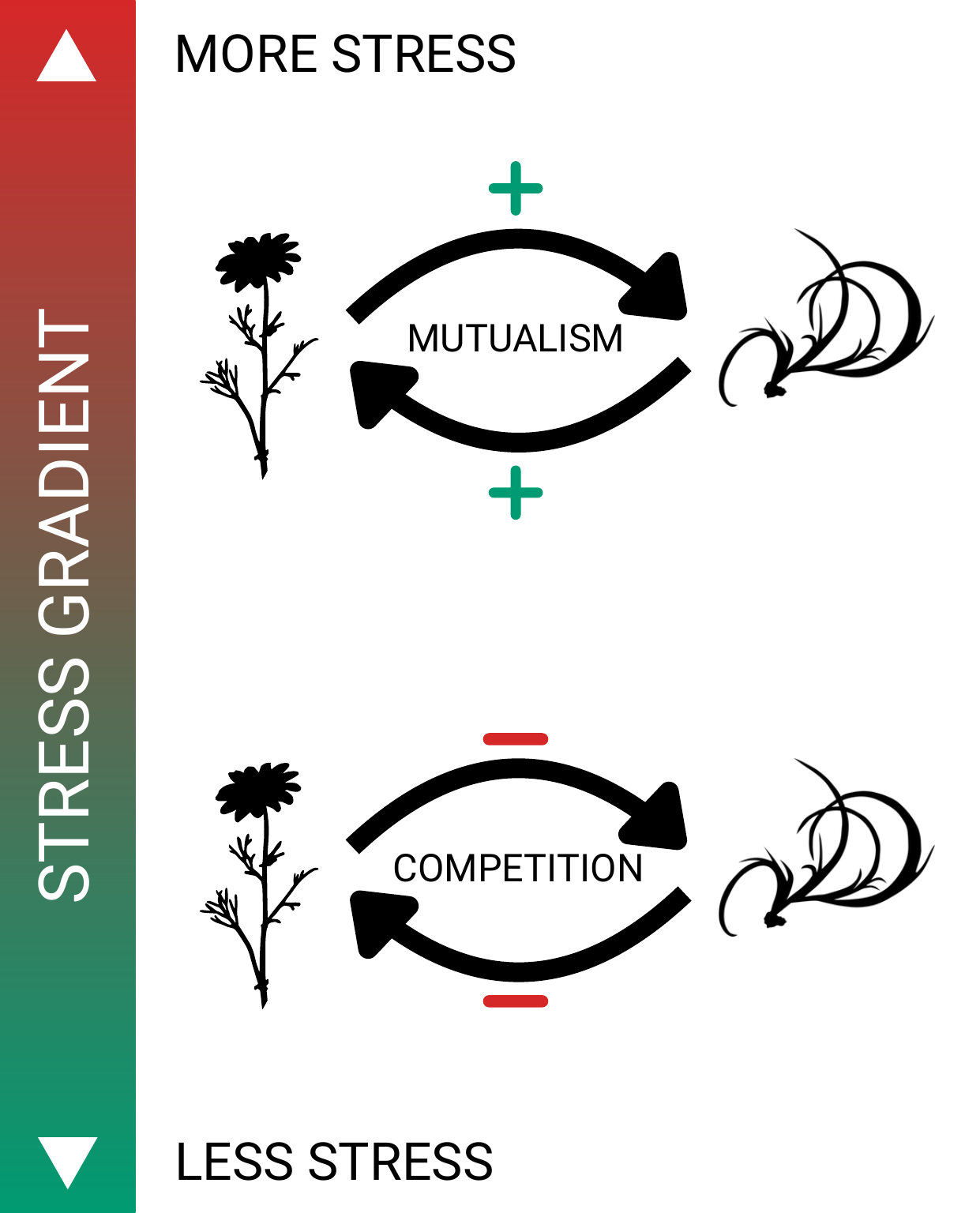}
            };
            \node at (-3.4,4.8) {\textbf{a)}};
        \end{tikzpicture}
    \end{minipage}
    \begin{minipage}[r]{0.55\linewidth}
        \centering
        \begin{tikzpicture}
        \node at (0.5,0) {
        \includegraphics[width=1.0\textwidth]{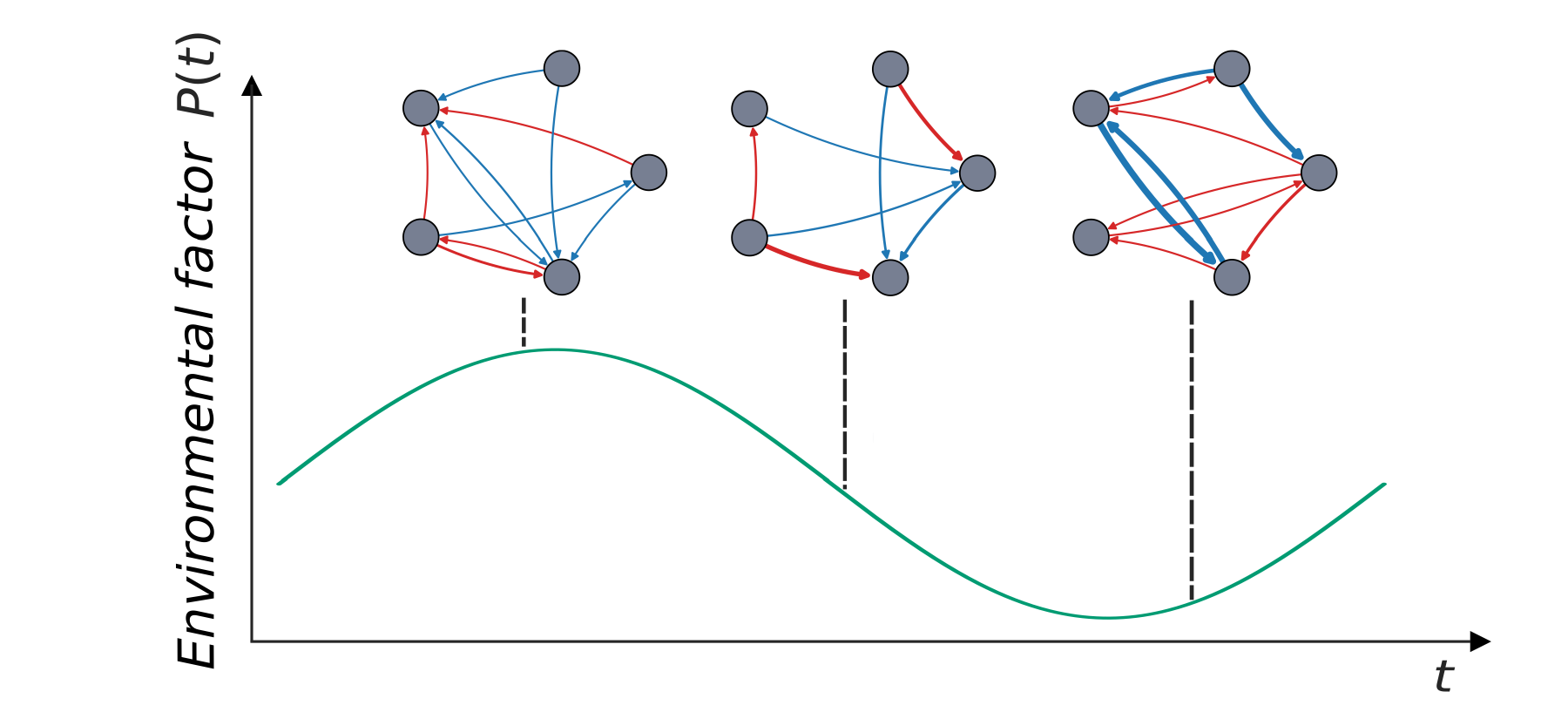}
        };
        \node at (-3.45,1.8) {\textbf{b)}};
        \end{tikzpicture}
        \begin{tikzpicture}
        \node at (0,0) {
        \includegraphics[width=0.9\textwidth]{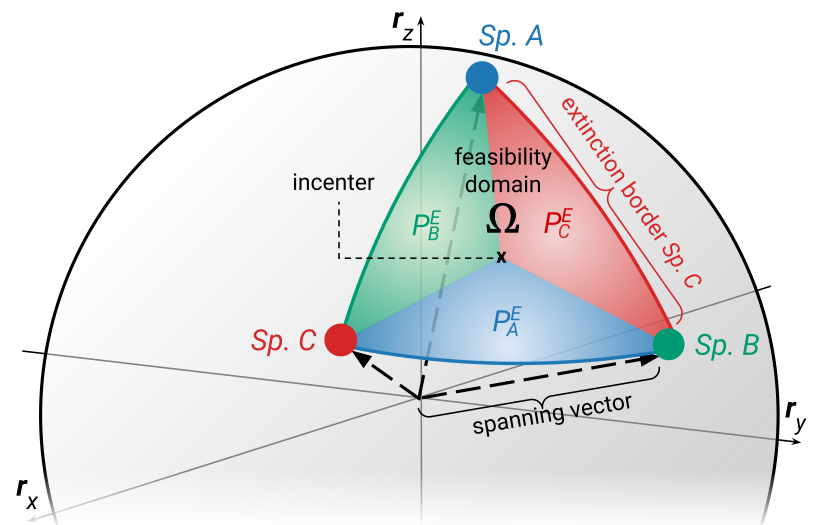}
        };
        \node at (-3.9,2.3) {\textbf{c)}};
        \end{tikzpicture}
    \end{minipage}
    \caption{
        \textbf{Ecological system characteristics.}
        \textbf{a)} The stress gradient hypothesis postulates that species became increasingly competitive under low environmental stress, and increasingly mutualistic under more environmental stress.
        \textbf{b)} Illustrative environmental factor as function of time and the respective representation of the temporal networks describing the wild bees system (\textit{BEEFUN}). As the environment changes, the interactions of the system changes in order to adapt to the new environment. That change is being captured by the appearance/disappearance of interactions with respect to the calculated threshold, flipped link signs, and interaction strength variations. For the wild bees system, the environmental factor that we are considering and that affects the dynamics of the system is the rainfall.
        \textbf{c)} Illustration of the feasibility domain ($\Omega$) of a three-species ($A$, $B$, $C$) interacting system and its corresponding elements. The outer limit of each region corresponds to a species' extinction border, i.e., the most extreme case in which it may survive before facing extinction-level events due to higher environmental stress.
    }
    \label{fig:1}
\end{figure}
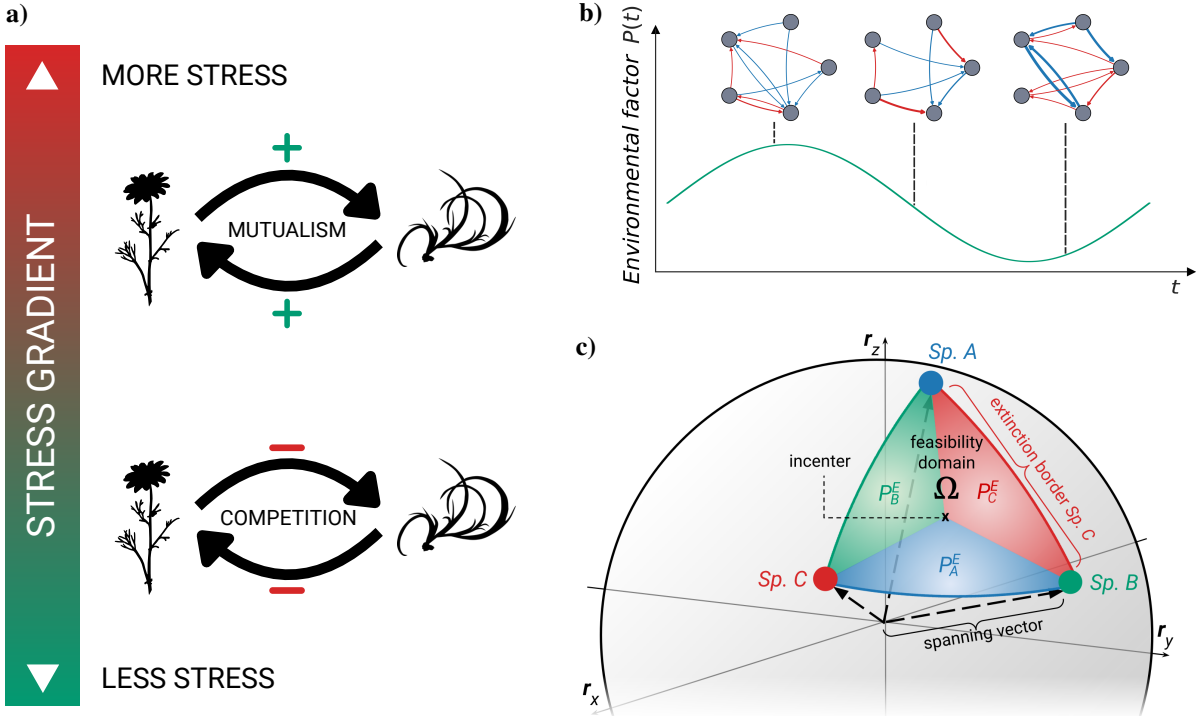

The study of ecological interactions is essential to understanding the complexity of ecological communities. Representing these interactions —such as predation, competition, cooperation, and parasitism— as networks has uncovered that they exhibit structural recurring patterns (such as modularity and nestedness \cite{bastolla_architecture_2009}). Their effect (i.e. negative, neutral, or positive) and their strength govern dynamics and coexistence \cite{abrams_resource_1980, kefi_network_2015, adler_competition_2018}. For example, the stress-gradient hypothesis (Fig.~\ref{fig:1}a) predicts that, as environmental stress intensifies (e.g., drought, salinity, shading), mutualistic (positive) interactions become more common, whereas under benign, low-stress conditions, competitive (negative) interactions dominate \cite{maestre_refining_2009}.

One side of the complexity of ecological communities is that they are in constant change under shifting environmental conditions, so we expect interactions to rewire, strengthen, weaken, or disappear in response to that variation \citep{caradonna_seeing_2021}, Fig.~\ref{fig:1}b. Indeed, there is growing evidence that even when overall network architecture appears stable, individual links often fluctuate dramatically over time \citep{cai_temporal_2024, ushio_fluctuating_2018}. Understanding how these interactions shift —and what mechanisms help communities persist through such changes— is an urgent challenge in ecology, especially as the current human-driven environmental change accelerates \citep{strona_environmental_2016, meyer_temporal_2024}.

However, the majority of ecological results regarding network structure and its influence on dynamics have been obtained assuming static interactions, do not vary through time. Then, the consequences of the temporal interaction variation at the community level, paired with environmental shifts, have not been empirically evaluated yet. This gap between observation and theory can be attributed to two main limitations: the lack of long-term data and a reliable procedure to infer time-varying interactions. Fortunately, growing recognition of the value of extended studies —and the resulting influx of new data— now allows to integrate long-term observations with approaches to infer species interactions. Yet a major difficulty of these approaches is that the number of parameters increases exponentially with the size of the community. Then, many studies have used proxies of interactions, such as co-occurrence patterns, to infer species interactions from joint distributions \citep{cai_temporal_2024,momal_tree-based_2020}, or directly random graph models as null models or theoretical surrogates for empirical networks, offering baseline expectations for network properties and dynamics \citep{landi_complexity_2018}. Other methods estimate interaction strengths by examining how species abundances change along environmental gradients \citep{shang_inferring_2017}. Each of these approaches has inherent limitations: co-occurrence approaches risk conflating correlation with causation; environmental gradient methods depend heavily on accurate environmental measurements; and random graph models lack the biological realism necessary to reflect true ecological dynamics. These limitations can introduce significant unpredictable biases in species interactions and their interpretation \citep{bluthgen_critical_2024}.

Here, we do a first step towards characterising how species interactions vary over time due to environmental shifts, thanks to five long-term time series of species abundances across multiple locations and basic but reliable network inference, Table~\ref{tab:ecological_networks}. Our systems experienced different levels of environmental regimes (rainfall, temperatures), and differed in size, and taxa, factors known to modulate context-dependent shifts in species performance, interactions, and persistence. To obtain temporal interaction networks, we parameterized population models for each system, proposing that species dynamics are linked to their environmental factor by time-varying interaction strengths, Fig.~\ref{fig:ratio-and-feasibility-domain}b. This framework revealed that the observed abundance trajectories are best captured by allowing interaction strengths to vary. With this information in hand, we are in a position to address three pressing needs: (i) quantify how environmental variability shapes interaction rewiring and strengthening; (ii) disentangle whether these temporal changes are at a stationary equilibrium; and (iii) clarify how the effect of environment shifts in the interactions influences coexistence.

To gain a mechanistic insight into how time-varying interactions influence coexistence, we embed our framework within the theory of structural stability \cite{saavedra_structural_2017, rohr_structural_2014, gonzalez-casado_evidence_2025}. In essence, structural stability assesses the range of conditions under which ecological communities remain feasible, that is, when all species present positive abundances. It holds that a community is structurally stable if the so-called Feasible Domain $\Omega$ accommodates each species’ performance asymmetries, Fig.~\ref{fig:1}c. Then, the size of $\Omega$ is a proxy for the coexistence opportunities of a community. Critically, the size of $\Omega$ is determined by the structure of species interactions. Therefore, our main expectation is that the temporal changes in the interaction network induced by environmental shifts shape the coexistence opportunities of communities. 

The exploration of the five long-term datasets and our time-varying framework produces a temporal series of interaction networks for which we do not assume their nature beforehand. The links are signed, weighted, and directed, and are in those values where the effect of the environment is encoded. Though these rich temporal networks are notoriously difficult to characterise, network scientists have risen to the challenge —developing innovative tools to capture their evolving structure \cite{bauza_mingueza_characterization_2023, gomez-ambrosi_measuring_2025}. In particular, our first step is describing the overall changes in the structure of species interactions, especially interaction rewiring, since phenotypic plasticity and behavioral flexibility are key drivers of species persistence and ecosystem functions \cite{turcotte_phenotypic_2016, godoy_excess_2020}. Building on this descriptive step, we then turn to the question of whether —despite continual change of individual links— the network’s macroscopic topology remains consistent, merely fluctuating within predictable bounds. To test this, we examine whether the probabilistic rules governing link dynamics (creation, destruction, and weight and sign shifts) are stationary over time \cite{gonzalez-casado_evidence_2025}. Demonstrating such stationarity is crucial: it would reveal a dynamic equilibrium among competing ecological processes and indicate a balance between resource availability and species’ functional demands. We finally examine the consequences of these environment-driven temporal changes in species interactions for the stability. To link temporal rewiring with community-level stability, we use structural stability—an approach that quantifies the range of demographic conditions allowing species to coexist, based on their interactions and performance \cite{saavedra_structural_2017}. 

Equipped with this battery of analyses, we can assess how environmental shifts translate into interaction changes and, ultimately, into patterns of community coexistence. In Section~\ref{sec:methods}, we introduce our temporal-variation metrics; Section~\ref{subsec:results_visualization} then demonstrates that interactions do change over time in ways that carry ecological meaning. Section~\ref{subsec:results_visualization_2} reveals that --- despite continual rewiring --- link transitions settle into a dynamic equilibrium.
Lastly, Section~\ref{sec:discussion} shows how these coordinated changes preserve --- or even strengthen --- community structural stability according to environmental harshness.

\section{Methods and Materials}
\label{sec:methods}

Interaction networks are constructed from the abundance timeseries of 5 long-term observational studies, spanning different taxa, numbers of species, and temporal lengths as detailed in Table~\ref{tab:ecological_networks}. Each study focused on the early influence of different environmental factors $P(t)$ on the populations. The obtained networks are signed, weighted, and directed.

\begin{table}[ht]
	\centering
	\caption{Summary of the ecological networks used in this study.}
	\begin{tabular}{llllll}
	 \textbf{Dataset} &\textbf{System} & \textbf{Species} & \textbf{Years} & \textbf{Environmental factor $\boldsymbol{P(t)}$} & \textbf{Ref.}\\
     \hline \\
	BEEFUN &  Wild Bees &5 & 8 & Rainfall & \cite{dominguez-garcia_interaction_2024}\\
      CARACOLES & Annual Plants & 7 & 9 & Rainfall &\cite{garciacallejas_nonrandom_2023}\\
	DIG\_13 & Seabirds in Fisheries & 3 & 43 & Sea surface temperature  & \cite{barbraud_density_2018}\\
	 DIG\_50 & Seabirds at Barents Sea & 3 & 27 & Temperature & \cite{durant_within_2012}\\
	LPI\_2858 & Lizards & 6 & 15 & Rainfall & \cite{read_booming_2012}\\
	\end{tabular}
	\label{tab:ecological_networks}
\end{table}


\subsection{Time-varying framework}

To parameterize empirically derived interaction estimates, we modelled changes in abundances using a Ricker model, which can be interpreted as a discrete-time formulation of the generalized Lotka–Volterra equations:

\begin{equation} \label{eq:ourmodel}
    \log \left(\frac{N_{i}(t+1) + 1}{N_{i}(t) + 1}\right) = 
    r_i + r_i'P(t) +
     \sum_{j = 1}^n \left(A_{ij} + B_{ij}  P(t) \right) N_{j}(t),
\end{equation}
where $n$ is the number of species in the community, $N_i(t)$ denotes the abundance of species $i$ at a time $t$; $r_i$ is the intrinsic growth rate of species $i$; and $r_i^\prime$ represents the effect of environment on it.  The term $A_{ij}$ refers to the interaction matrix describing the effects between species that do not depend on the environment, while $B_{ij}$ captures the influence of the environmental factor on each species interaction. Lastly, $P(t)$ represents the value of the environmental factor at time $t$.

Then, the time-varying species responses as an effect of the environmental conditions at each time $P(t)$, in both intrinsic growth rates and interactions, were introduced as:

\begin{equation}
     \tilde{r}_i(t) = r_i + r_i'P(t) ,
\end{equation}
 \begin{equation}\label{eq:A_t_model_1}
     \tilde{A}_{ij}(t) = A_{ij} +  B_{ij} P(t).
 \end{equation}

This choice was statistically more strongly supported and offered a better fit (lower AIC) and prediction than the model without the temporal variation and other time-dependent models. We fitted the parameters using the R package nlme for generalized linear-mixed models. Eq.~\ref{eq:A_t_model_1} represents the most general case and is the one used for two of our datasets: \textit{BEEFUN} and \textit{CARACOLES}, which involve annual solitary bees and annual plants. They offered the most complete fit because were sampled across multiple sites, resulting in more than 250 hours of field observations \cite{dominguez-garcia_interaction_2024} and 324 sampling plots \cite{garciacallejas_nonrandom_2023}, respectively. On the contrary, the other datasets concerning seabirds (\textit{DIG\_13} and\textit{ DIG\_50}) and lizards (\textit{ LPI\_2858}) offered one sample per year, and thus it is not possible to infer the matrix $A_{ij}$ directly from the data. In such situations, a simplified version of the model is used, where the effective parameters are given by
\begin{equation}\label{eq:r_model_2}
    \tilde{r}_i = r_i, 
\end{equation}
\begin{equation}\label{eq:A_t_model_2}
    \tilde{A}_{ij}(t) = B_{ij} P(t).
\end{equation} 
Note that this strong assumption precludes interspecies interactions from changing sign, condensing environmental effects on the variation of interaction strengths.



\subsection{Characterizing temporal changes}\label{subsec:links_changes}


We describe the changes in the interspecific interaction matrices by computing the number of sign changes, as well as link appearances and disappearances, over time. 
To identify ecologically meaningful interspecific interactions in this regard while accounting for the variability of interaction strengths, we apply an asymmetric thresholding procedure to each interaction matrix $\tilde{A}_{ij}(t)$. This procedure distinguishes between positive and negative interactions, defining separate thresholds for each type, based on the variability of the respective distributions:

\begin{equation}
    \sigma_+ = \text{std}(\tilde{A}_{ij}(t) \mid \tilde{A}_{ij}(t) > 0,\ i \neq j\}) \quad \text{,} \quad
    \sigma_- = \text{std}(\tilde{A}_{ij}(t) \mid \tilde{A}_{ij}(t) < 0,\ i \neq j\}.
\end{equation}    

At each time step $t$, an interspecies interaction is thus retained if

\begin{equation}
\left( \tilde{A}_{ij}(t) > 0 \ \text{and} \ \tilde{A}_{ij}(t) > \frac{ \sigma_+}{2} \right)
\quad \text{or} \quad
\left( \tilde{A}_{ij}(t) < 0 \ \text{and} \ \tilde{A}_{ij}(t)| > \frac{ \sigma_-}{2} \right).
\end{equation}

This asymmetric approach allows to maintain the strongest positive and negative interactions relative to their respective distributions, ensuring that the threshold used is adapted to the statistical structure of the matrix at each time step. All retained interactions are stored in a signed, directed, and weighted graph $\mathcal{G}(t)$. 

This type of analysis can be performed for \textit{BEEFUN} and \textit{CARACOLES} datasets. In these cases, the model allows to observe actual changes in the sign or presence of individual links over time, since it presents a constant matrix $\mathbf{A}$ that receives contributions through the modulation introduced by the term $\mathbf{B} \cdot P(t)$, as stated in Eq.~\ref{eq:A_t_model_1}. If $A$ is a zero matrix --- as in the cases of \textit{LPI\_2858}, \textit{DIG\_13}, and \textit{DIG\_50}, Eq.~\ref{eq:A_t_model_2} --- and the environmental parameter $P(t)$ is always positive, then the sign of the links in $\tilde{A}_{ij}(t)$ remains the same as in $B_{ij}$. In this case, the role of the environment is solely to modulate the intensity of the interactions over time.




\paragraph{Cooperation-competition ratio of interspecific interactions:}
One approach to quantifying the balance between cooperative (positive) and competitive (negative) interspecific interactions in a signed weighted ecological network is to compute the ratio of positive to negative interaction strengths. This can be achieved by summing all positively weighted interactions $\mathbf{\tilde{A}}^{+}$ and dividing this value by the sum of all negatively weighted interactions $\mathbf{\tilde{A}}^{-}$:
\begin{equation}
    R(t) = \frac{\sum \tilde{A}^{+}_{ij}(t)} { \sum |\tilde{A}^{-}_{ij}(t)|},
\end{equation}

A ratio $R>1$ indicates that the cumulative strength of cooperative interactions outweighs that of competitive ones. Conversely, a value of $R=1$ suggests a balance between cooperation and competition, where neither interaction type predominates.

\paragraph{Equilibrium in the Interaction Networks:}

To assess whether the networks under study are at equilibrium, we follow the procedure described in \cite{gonzalez-casado_evidence_2025}. Because the edges in our networks represent species interactions and can take any continuous value, we first discretize them to facilitate analysis of interaction dynamics over time. We adopt a simple and interpretable discretization into two categories: positive interactions (encoded as 1) and negative interactions (encoded as 0). This results in four possible interaction types between species pairs: \texttt{00}, \texttt{01}, \texttt{10}, and \texttt{11}. Consequently, there are $4^2$ possible transitions between interaction types across two consecutive time points.

For each pair of consecutive time observations $K$, we compute a transition probability matrix $m^K$ by counting the frequency of each of the 16 possible transitions and normalizing the counts by row. This yields the empirical transition probabilities of interaction types between time steps.

To determine whether the system has reached stationarity, we test for the statistical equivalence of transition matrices over time. Instead of performing exhaustive pairwise comparisons, we compare each $m^K$ to the average transition matrix $\langle m \rangle$. This approach allows to assess temporal stability while reducing the number of comparisons \cite{gonzalez-casado_evidence_2025}.

Finally, to evaluate whether the system is at equilibrium, we verify two conditions for each transition matrix: the system has reached a stationary state, and the detailed balance condition is satisfied. 

\subsection{Structural Stability}
\label{subsec:feasibility_domain}

To address how the temporal nature of interactions shapes the opportunities of coexistence, we investigate three different quantities: the predicted feasible solution for the communities at each environmental value $P$ ($\boldsymbol{N}^*_P$), the size of the Feasibility Domain $\Omega$, and the asymmetry parameter $J(\boldsymbol{\tilde{A}})$, whose interpretation will be explained shortly.

Lets assume that for a given time $t^*$, the system population vector $\boldsymbol{N}(t) \in \mathbb{R}^{n}$ can reach a stationary state, {\itshape{i.e.}}, $\boldsymbol{N}(t)=\boldsymbol{N}(t') =\boldsymbol{N}^*_P, \, \forall \,t,t'\geq t^*$. The subindex $P$ represents the fact that the stationary solutions are conditioned to the value of the environmental factor at which the system has hypothetically thermalized. In this regime, the left-hand side of Eq.~\ref{eq:ourmodel} is equal to zero, yielding
\begin{equation}\label{Nf_thermal}
    \boldsymbol{N}_P^*(\boldsymbol{\tilde{r}} - \mathbf{\tilde{A}} \boldsymbol{N}^*_P) = 0,
\end{equation}
where $\boldsymbol{\tilde{r}}$ and $\mathbf{\tilde{A}}$ are the effective stationary intrinsic growth rates and adjacency matrix. The non-trivial solution of the previous equation, obtained by imposing that all components of $\boldsymbol{N}^*_P$ are positive, defines the feasible population vector as
\begin{equation}\label{Nf_model}
    \boldsymbol{N}_P^* = -\mathbf{\tilde{A}}^{-1} \boldsymbol{\tilde{r}}.
\end{equation}
Notice that, when supplied with empirical intrinsic growth rates and adjacency matrices, Eq.~\ref{Nf_model} may present a seemingly pathological behavior of negative populations. Such behavior makes no ecological sense, and it should be interpreted in the sense that our hypothesis (the community being feasible) is not valid for the specific data under consideration.

The set $D_F(\mathbf{A})$ of all possible $\boldsymbol{\tilde{r}}$ that allow for non-trivial feasible solutions of Eq.~\ref{Nf_thermal} is called the Feasibility Domain \cite{saavedra_structural_2017}. Geometrically, it can be interpreted as the surface containing all $\boldsymbol{\tilde{r}}$ that can be written as a linear combination of positive numbers ($\boldsymbol{N}^*_P$) of (minus) the column vectors of the effective interaction matrix $A_i$, denoting as "spanning vectors" in Fig.~\ref{fig:1}c:
\begin{equation}\label{rFD}
    \boldsymbol{\tilde{r}} = \{ N^*_{1,P}(-A_1) + \cdots + N^*_{n,P}(-A_n) \}.
\end{equation}
To infer the size of the feasibility domain, we usually project these spanning vectors over the (hyper)sphere of radius 1 (see Fig.~\ref{fig:1}c for an example). Denoting by $B_S$ the area of the unitary radius $n$-dimensional hypersphere, the size of $\Omega$ is formally given by \cite{allenperkins_structural_2023}:
\begin{equation}\label{Omega}
    \Omega(\mathbf{\tilde{A}}) = \frac{\text{vol}(D_F(\mathbf{\tilde{A}}) \cap B_S)}{\text{vol}(B_S)}.
\end{equation}
It is a direct measure of the probability that all species persist, indicating the ecosystem's overall tolerance to environmental variations, and it can be computed using a quasi-Monte Carlo method in the R package \texttt{anisoFun}~\cite{allen-perkins_radicalcommecolanisofun_2023}.

To gain a deeper understanding of the structural stability, other geometrical aspects of the Feasibility Domain can be considered. For a given size, the shape of the Feasibility Domain reveals the relative vulnerabilities of individual species to isotropic perturbations \cite{allenperkins_structural_2023}. To quantify these vulnerabilities, we can compute the probability $P^E_i( \mathbf{\tilde{A}}) $ that species $i$ will be the first to be excluded if a perturbation in a random direction occurs. In this way, we estimate the fraction of $D_F(\mathbf{\tilde{A}})$ where $\boldsymbol{\tilde{r}}$ are closer to the link where species $i$ is excluded, represented by colored areas in Fig.~\ref{fig:1}. Following Eq.~\ref{Omega}, we can write the probability as a volume ratio:

\begin{equation}
P^E_i( \mathbf{\tilde{A}}) = \frac{\Omega({\tilde{A}_i^M})}{\Omega(\mathbf{{\tilde A}})},
\end{equation}

where $\mathbf{{A_i^M}}$ is a modified interaction matrix where the $i$-th column vector of the matrix has been replaced by the incenter vector of the feasibility domain, corresponding to the location in which the minimum distance to any of its borders is the same regardless of the direction of the perturbation.

To quantify the asymmetry in species vulnerabilities and thus to qualify the shape of the feasibility domain, we use the asymmetry index $J'$ \cite{allenperkins_structural_2023}, based on the relative Shannon diversity index \cite{pielou_ecological_1975}. It yields:

\begin{equation}
    J(\mathbf{\tilde{A}} ) = -\frac{\sum_{i=1}^n P^E_i(\mathbf{\tilde{A}}) \log P^E_i(\mathbf{\tilde{A}})}{\log(n)}.
\end{equation}

This index possesses three key properties that make it particularly suitable for ecological analysis: (i) its maximum value of one corresponds to the case where all species have equal exclusion probabilities of (symmetric feasibility domain); (ii) it approaches zero asymptotically when exclusion probabilities are highly inhomogeneous among species; and (iii) it shows intermediate values when survival differences between species can be considered moderate.

The correlation between the feasibility vector components and the environmental value at which it thermalized may be a useful quantity to combine with the entropic information contained in the asymmetry index, as it shows these effects from the point of view of population densities.

\section{Results}
\label{sec:results}

We begin with the analysis of link variability in order to have a general overview of the characteristics that each dataset and fitting model presented. To show the most important results, we selected both \textit{CARACOLES} and \textit{BEEFUN} datasets, which are described by the most general model in which interactions are described as a baseline interaction $A_{ij}$ plus an environmental effect $B_{ij} P(t)$; whereas the other datasets, are described by the more constrained version of the model in which $\tilde{A}_{ij} = B_{ij} P(t)$.

\subsection{Characterising time-varying ecological networks} \label{subsec:results_visualization}

\begin{figure}[htbp]
    \centering
    \includegraphics[width=1.1\textwidth]{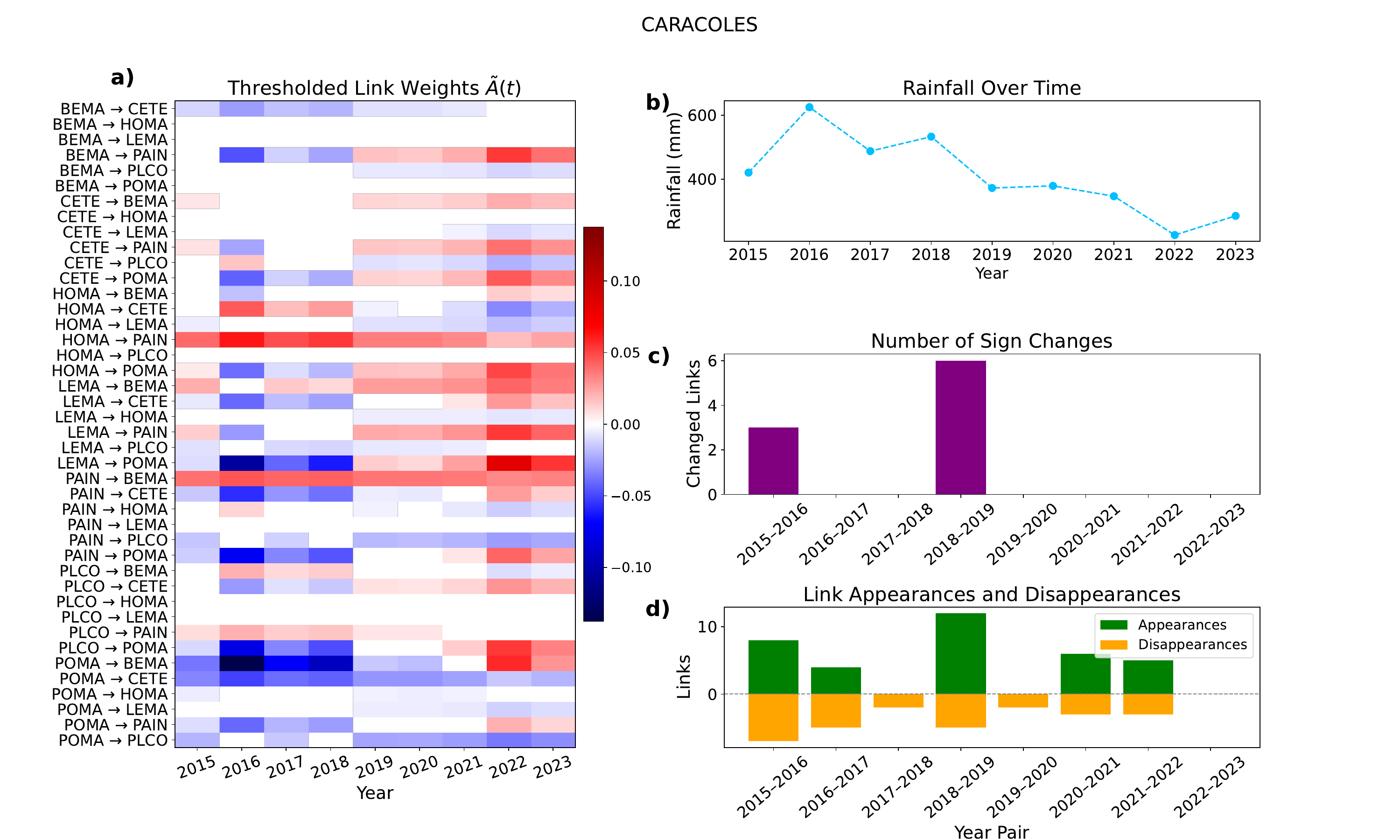}
    \caption{\textbf{Link variability analysis for the \textit{CARACOLES} dataset.} \textbf{a)} Heatmap showing the thresholded link weights over time for all possible pairwise interactions. All pairwise interactions are represented along the $y$ axis. For this dataset the link weights change their sign and their intensity over time. In order of appearance: \textit{Beta macrocarpa} (BEMA), \textit{Centarium teniuflorum} (CETE), \textit{Hordeum maritimim} (HOMA), \textit{Leontodon maroccanus} (LEMA), \textit{Parapholis incurva} (PAIN) and \textit{Plantago coronopus} (PLCO). \textbf{b)} Rainfall values as a function of time. Rainfall is the environmental factor associated with the ecological system described by dataset \textit{CARACOLES}. \textbf{c)} Number of sign changes within the interaction networks in two consecutive years. There are two peaks of sign changes in these networks, one in the transition from 2015 to 2016, and a higher one in the transition from 2018 to 2019. \textbf{d)} Appearance/disappearance of links within the interaction networks in two consecutive years. Links are appearing in every time step, with an exception in the transitions from 2017 to 2018 and 2019 to 2020. There are links disappearing in every time step.}
    \label{fig:CARACOLES_links}
\end{figure}

\begin{figure}[htbp]
    \centering
    \includegraphics[width=\textwidth]{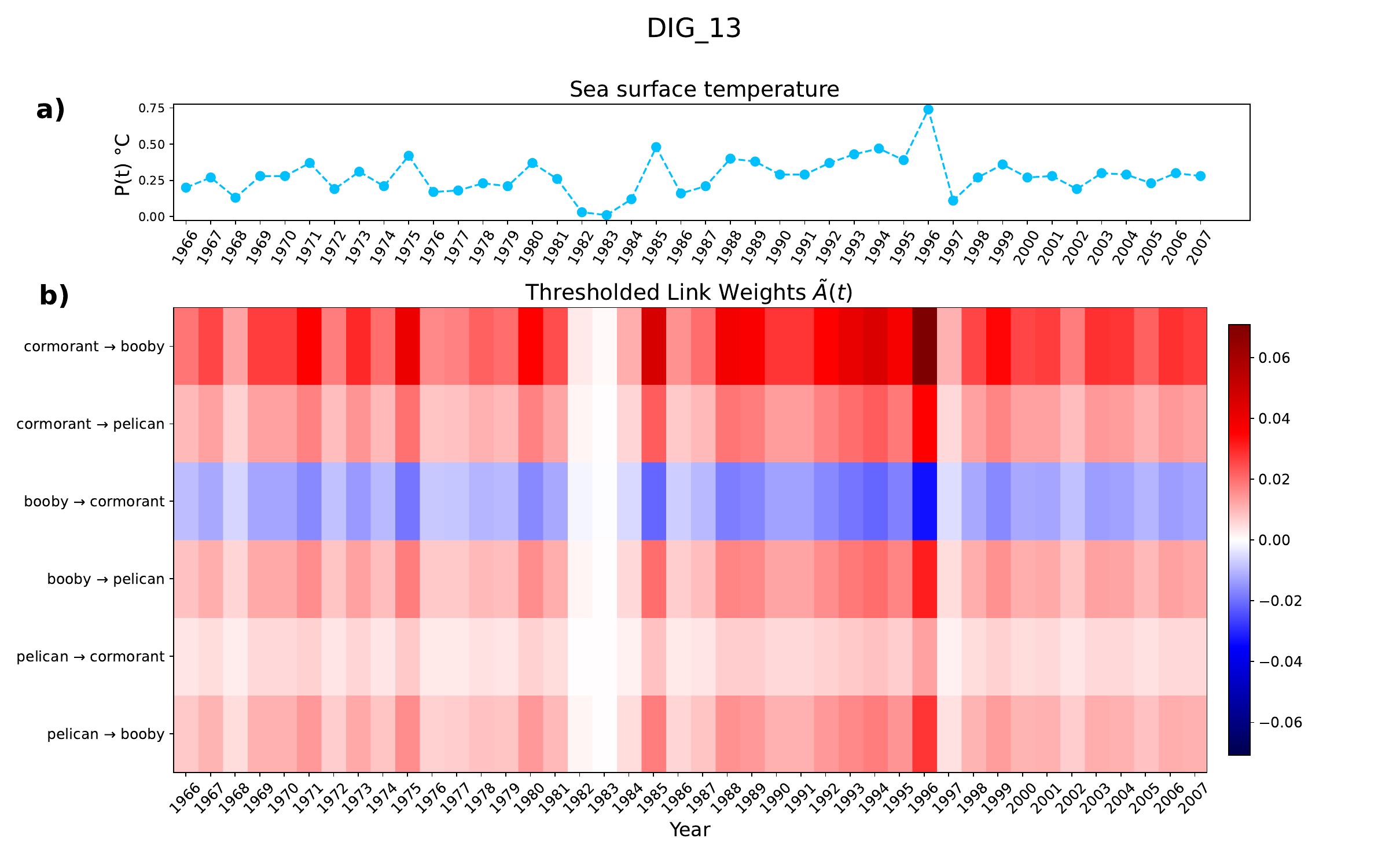}
    \caption{\textbf{Link variability analysis for the \textit{DIG\_13} dataset.} \textbf{a)} Sea surface temperature as a function of time, the environmental factor associated with the ecological community of seabirds described by dataset \textit{DIG\_13}. \textbf{b)} Heatmap showing the links over time for all possible pairwise interactions. For this dataset, the links can not change their sign, but can and do vary in intensity according to the environmental factor (i.e., by the matrix $\mathbf{B}$).}
    \label{fig:DIG_13_links}
\end{figure}

Regarding the results obtained for the \textit{CARACOLES} dataset, Fig.~\ref{fig:CARACOLES_links} shows links that vary in sign and rewire (appear and disappear) when the environmental factors change. That observation demonstrates that the model is able to capture the system’s readaptation following an environmental shock. For example, from 2015 to 2016, we observed a sharp increase in rainfall, which corresponds to a peak in sign changes as well as in link appearances and disappearances Fig.~\ref{fig:CARACOLES_links}b and ~c. A similar effect is observed during the drop in rainfall from 2018 to 2019, where the peaks in sign changes and link dynamics are even more pronounced.
The remaining datasets are analysed in the Supplementary Information (Figs.~S3, ~S4 and ~S5). In the datasets where sign changes could not be modelled due to fitting constraints (as \textit{DIG\_13}), the values of the environmental effects are not able to produce temporal changes other than strengthen some weights, with some links being affected more by environmental shifts as highlighted in Fig.~\ref{fig:DIG_13_links}b.

Next, we performed the equilibrium analysis on the rich \textit{BEEFUN} and \textit{CARACOLES} datasets. For both datasets, the corresponding seven and eight transition matrices, respectively, were found to be statistically equivalent, indicating stationary dynamics. However, a few p-values fell below the significance threshold. In the \textit{BEEFUN} dataset, the proportions of transitions with significant differences were 0.0625, 0.125, and 0.0625 at the first, second, and fourth time intervals, respectively. In the \textit{CARACOLES} dataset, significant deviations occurred at the first and fourth time intervals, with corresponding fractions of 0.1875 and 0.3125. Given the low magnitude and isolated nature of these deviations relative to the total number of transitions, we attribute them to random fluctuations. We may thus conclude that the stochastic processes governing network evolution in both datasets are stationary, with consistent transition probabilities over time.

Furthermore, we found that the detailed balance condition for equilibrium was consistently satisfied for both the \textit{BEEFUN} and \textit{CARACOLES} datasets, at every observation time. This indicates that the probability flow between each pair of interaction states is symmetric, thus indicating that the networks derived from both datasets are at equilibrium.

\subsection{Characterising the temporal structure in terms of environmental changes}\label{subsec:results_visualization_2}

\begin{figure}[ht]
    \centering
    \includegraphics[width=\linewidth]{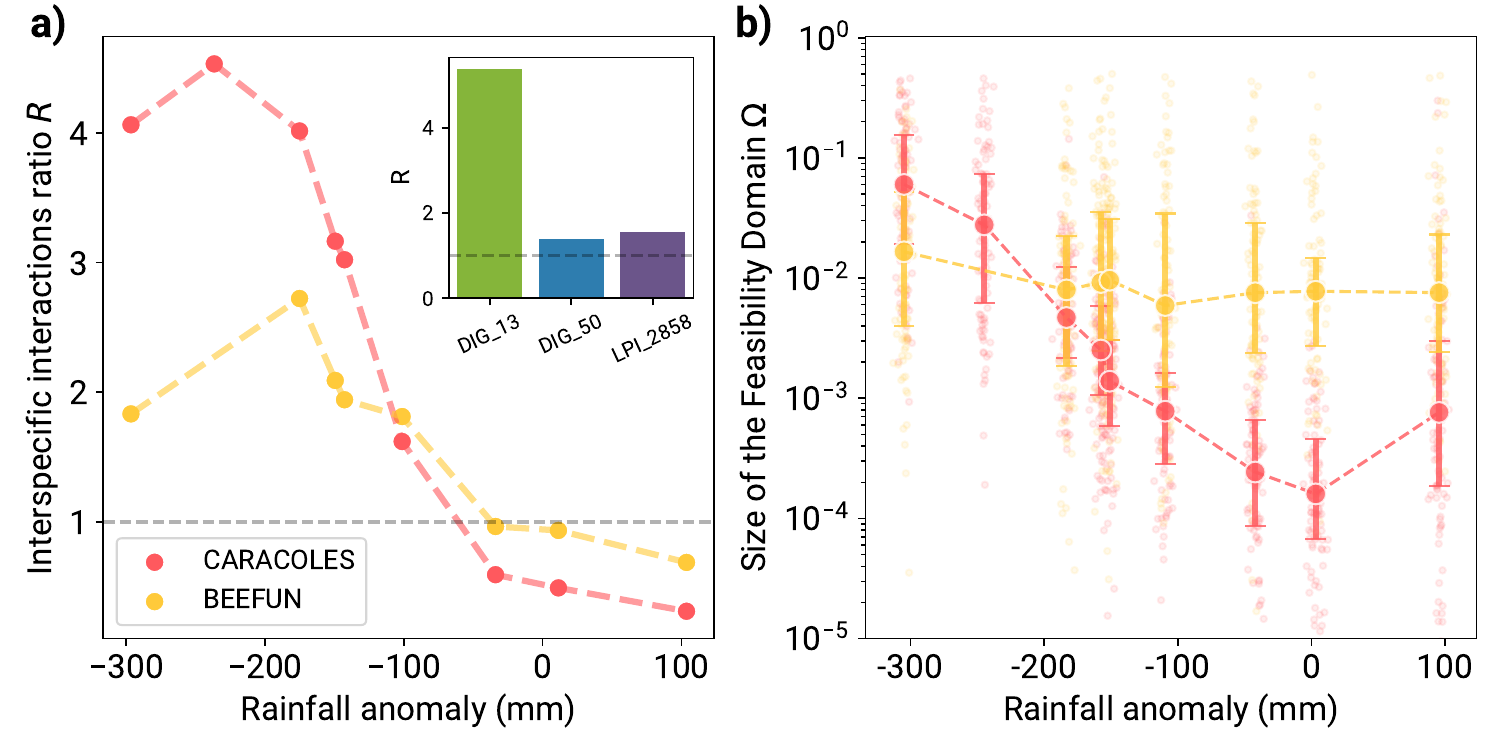}
    \caption{ \textbf{Impact of the environmental factor}. \textbf{a)} Cooperation-competition ratio as a function of the environmental factor (rainfall anomaly) for the networks in each dataset. On the inset the \textit{DIG\_13}, \textit{DIG\_50}, \textit{LPI\_2858} ratio of interspecific interactions, that are constant through environmental changes. For the \textit{BEEFUN} and \textit{CARACOLES} datasets, the interspecific interactions ratio shows a negative correlation with the environmental factor, with both starting in a cooperative state and then evolving to a competitive state as the environmental factor increases. For the \textit{{DIG\_13}}, \textit{{DIG\_50}} and \textit{{LPI\_2858}} datasets, the constant ratio shows a dominance of cooperative states.
    \textbf{b)} Evolution of the Feasibility Domain ($\Omega$) as a function of the environmental factor (rainfall anomaly). Bootstrap simulations are shown as points with reduced color intensity, while darker points represent the FD median for each environmental condition. For \textit{BEEFUN}, the FD remains constant, whereas for \textit{CARACOLES}, a negative correlation is observed.}
    \label{fig:ratio-and-feasibility-domain}
\end{figure}

So far, we have compared the changes in the structure of interaction networks between consecutive times. From an ecological point of view, the tendencies of the interactions with environmental variation may inform about the direction in which the environment pushes pairs of species. To quantify that, we tracked the values of interactions across the environmental gradients and calculated the cooperation-to-competition ratio of interspecific interactions. The \textit{BEEFUN} and \textit{CARACOLES} datasets exhibited a nonlinear correlation between the ratio and their environmental factor, expressed as an anomaly to make clearer the environmental extremes, Fig.~\ref{fig:ratio-and-feasibility-domain}a. The ratio decreases as the rainfall increases, indicating a transition in the network from a facilitative to a competitive interaction. Such a result aligns with the idea of the stress gradient hypothesis~\cite{bertness_physical_1994, maestre_refining_2009}, as the environment becomes harsh (drought in this case) positive interactions are favored ($R>1$). When environmental stress is released, competition interactions increase with respect to cooperation and the interspecific interactions ratio started to show a dominance for competitive interactions ($R < 1$).

In contrast, the \textit{DIG\_13}, \textit{DIG\_50}, and \textit{LPI\_2858} datasets have a constant cooperation-competition ratio as the environmental factor changes imposed by the model structure. In these networks, there is no change in the sign of interaction and the weights in each time are just a composition of the constant matrix $B_{ij}$ by the multiplicative factor $P(t)$, Eq.~\ref{eq:A_t_model_2}. However, looking at the values of $B_{ij}P(t)$ we see these communities are in a cooperative-dominated state ($R>1$), inset of Fig.~\ref{fig:ratio-and-feasibility-domain}a. Among the three, {DIG\_13} exhibited the highest ratio $R$, with cooperative interactions being approximately five times as prevalent as competitive ones. The other two datasets, \textit{DIG\_50} and \textit{LPI\_2858}, displayed very similar ratios of $R \approx 1.5$.

To gain an overall picture of the influence of environmental shifts on coexistence opportunities, we computed the size of the Feasibility Domain $\Omega$ (see Sec. S1 in the Supplementary Information) for \textit{CARACOLES}  and \textit{BEEFUN} (Fig.~\ref{fig:ratio-and-feasibility-domain}b)  datasets. In the latter, the size of the Feasibility Domain remains constant, despite the variation in rainfall, indicating that there are no significant changes in the region of coexistence of species for that ecological system (Fig.~S1). However, in the case of \textit{CARACOLES} dataset, we observed a decrease in size with increasing rainfall (Fig.~S2). This is an indication of the excess rainfall reducing the variability in intrinsic growth rates that is necessary for the long-term sustainability of this community.

\begin{figure}[ht]
    \centering
    \includegraphics[width=\linewidth]{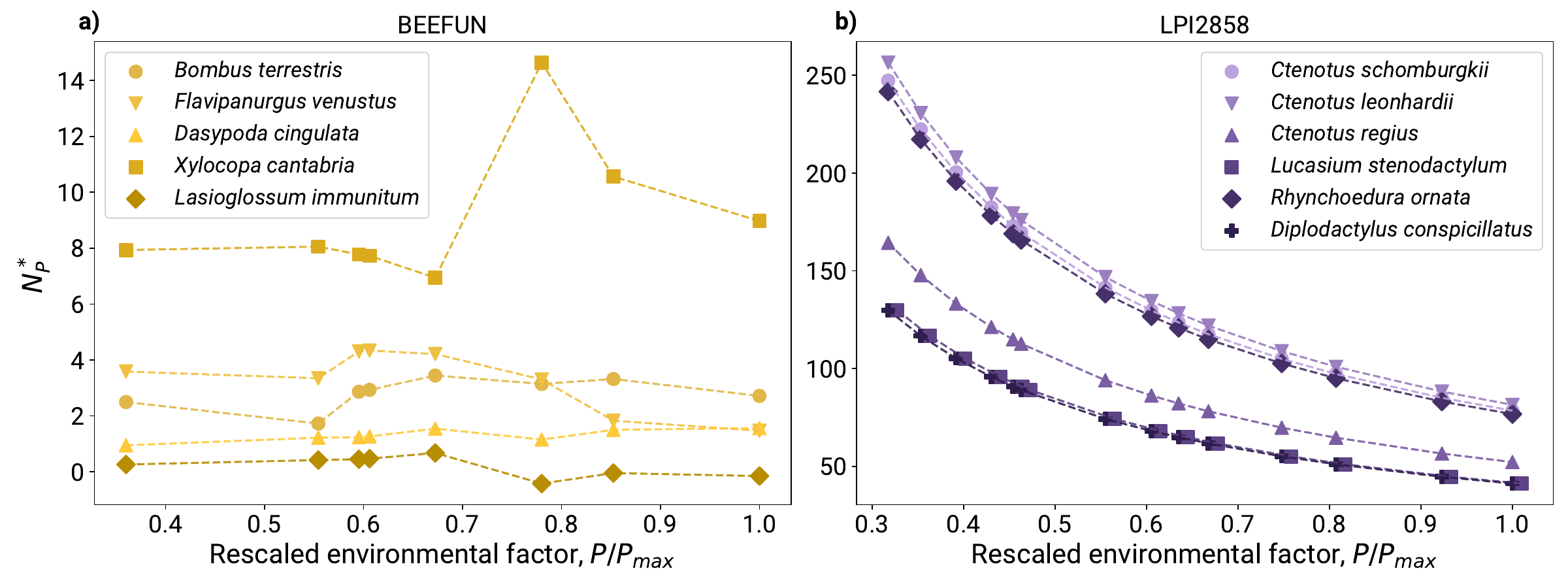}
    \caption{{\bfseries{Environmental effect on species abundance.}} Correlation between feasible species abundance and environmental stress for \textbf{a)} \textit{BEEFUN} and \textbf{b)} \textit{LSI\_2858} datasets. Equivalent plots for the {\textit{DIG\_13}}, {\textit{DIG\_50}} and {\textit{CARACOLES}} datasets are presented in the supplemental material. For the \textit{BEEFUN} dataset, the feasible species abundance shows some fluctuations with a non defined trend. For the \textit{LSI\_2858} dataset, the feasible species abundance decreases with environmental factor for all the species with a similar behavior, but different intensities.} 
    \label{fig:leo-NxP}
\end{figure}

Turning our focus to the values of each species abundances, in Fig.~\ref{fig:leo-NxP}, we present the correlation between the feasible species abundances and the environmental factors, for the {\itshape{BEEFUN}} (\ref{fig:leo-NxP}a) and {\itshape{LPI\_2858}} (\ref{fig:leo-NxP}b) datasets. The components of $\boldsymbol{N}^*_P$ were obtained through the solution of Eq.~\ref{Nf_model} for each of the environmental factors $P$. For {\itshape{BEEFUN}}, fitted using the complete effective adjacency matrix in Eq.~\ref{eq:A_t_model_1}, there is no evident relation between species abundance and environmental harshness, indicating that species intrinsic growth rates also play a role in the theorised final abundances for each time step. On the contrary, for {\itshape{LPI\_2858}}, fitted with the effective matrix in Eq.~\ref{eq:A_t_model_2}, it is shown that species abundance is inversely proportional to the environmental factors in which the system is assumed to have equilibrated. The latter supports that feasible abundances decrease as a response to an increase on the environmental-stress increases.  

\begin{figure}[ht]
    \centering
    \includegraphics[width=0.9\linewidth]{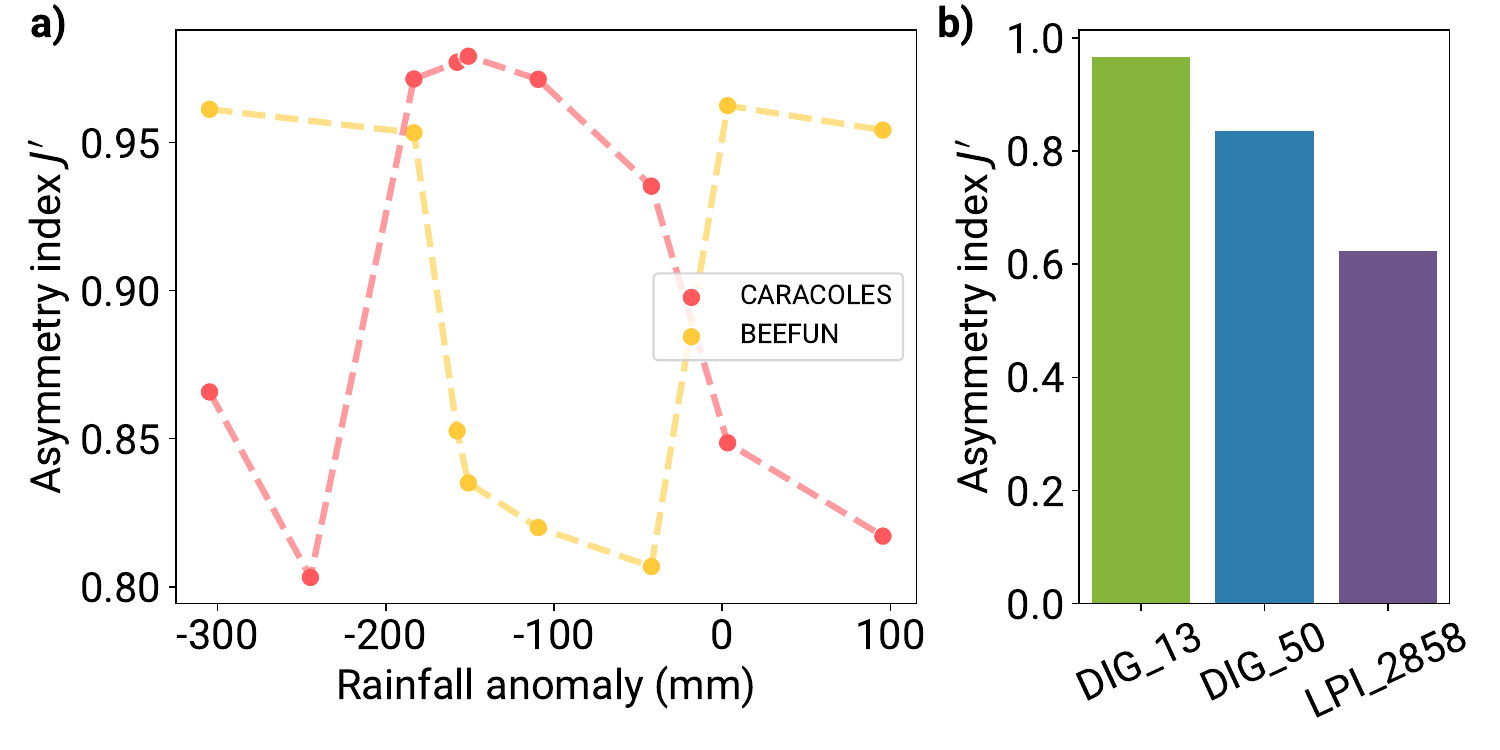}
    \caption{{\bfseries{Environmental effect on the asymmetry index.}} \textbf{a)} Correlation between asymmetry indexes $J^\prime$ and environmental factors for the \textit{BEEFUN} and \textit{CARACOLES} datasets. The greater the asymmetry index is more equally distributed is the FD among the species. Small values of the asymmetry index indicates that some species dominates the FD. The asymmetry index in both datasets showed a non trivial behaviour. \textbf{b)} Constant asymmetry indexes for the \textit{DIG\_13}, \textit{DIG\_50}, and \textit{LPI\_2858} datasets. \textit{DIG\_13} demonstrated the highest asymmetry index, indicating an homogenous distribution of the exclusion probabilities. In contrast, \textit{LPI\_2858} dataset showed the lowest asymmetry index, indicating a high inhomogeneous distribution of the exclusion probabilities among the species of that system.}
    \label{fig:asymmetry-index}
\end{figure}

In addition to the size of the Feasibility Domain, it is important to consider that its shape can also provide meaningful insights into the vulnerabilities of species within each ecological system. The size of the FD ($\Omega$ in Fig.~\ref{fig:1}) is a proxy of the tolerance of a community to random variations in their species performances (for a particular set of interactions). However, two communities with the same number of species $n$ and identical feasibility domain sizes can still present very different responses, depending on the shape of their feasibility domain. From this perspective, we calculated the asymmetry index of the feasibility domains at each environment value). The results for all datasets are shown in Fig.~\ref{fig:asymmetry-index}. As expected, we find that the \textit{DIG\_13}, \textit{DIG\_50}, and \textit{LPI\_2858} indexes remain constant as the environmental factor changes. Although, for the \textit{BEEFUN} and \textit{CARACOLES} datasets, the results demonstrated mirrored behaviors with the rainfall anomaly: one exhibiting a U-shaped curve and the other an inverted U-shaped curve. 

\section{Discussion}\label{sec:discussion}

While the prevailing paradigm in mathematical ecology assumes that interactions can be represented through static networks, our analysis underscores the importance of capturing the temporal dynamics of interacting species, as they reorganize themselves in response to changes in the environment. Our findings underline how species interactions adjust under changing environmental conditions over time, and we intend to address the limitations found in future work.

Differences in model complexity across the analysed datasets, and hence results, primarily stem from the availability of high-resolution temporal data. The temporal resolution with which ecological networks are constructed strongly influences the types of dynamic behaviors we can detect, as illustrated in Fig.~\ref{fig:CARACOLES_links} and Fig.~\ref{fig:DIG_13_links}. In particular, estimating only the environmental effect on interactions $B_{ij}$ without including a baseline interaction structure $A_{ij}$ cannot reveal critical dynamics, such as shifts in the sign of interactions.


The \textit{BEEFUN} and \textit{CARACOLES} datasets, which provided high-resolution temporal data across multiple locations, allowed us to fit the full model structure (Eq.~\ref{eq:A_t_model_1}). In these cases, our results (Fig.~\ref{fig:CARACOLES_links}) show that the main changes in interaction structure occurred in response to abrupt environmental shifts, confirming the sensitivity of interaction networks to external drivers. In contrast, the \textit{DIG\_13}, \textit{DIG\_50}, and \textit{LPI\_2858} datasets lacked sufficient temporal resolution. This constraint limited our ability to detect key patterns, such as sign flips.

We note that the need to resort to a simpler modelling framework in these cases emphasizes the critical role of temporal resolution in capturing the dynamic nature of ecological communities.
Most importantly, our analysis has focused mainly on interspecific interactions, while we leave an in-depth characterization of other mechanisms that involve intraspecific interactions (self-limitation) for future work.

\subsection{Equilibrium in Time-Evolving Networks}
The equilibrium analysis indicates that both the \textit{BEEFUN} and \textit{CARACOLES} datasets are governed by stationary dynamics and satisfy the conditions for equilibrium. This result highlights the value of using transition matrices to describe inherently non-equilibrium systems, as ecological communities. Among the mechanisms driving this equilibrium can be the tendency of species to adapt their interactions in response to environmental changes, yet only within limits imposed by resource availability and ecological roles. These constraints may restrict the extent of interaction variability, promoting relatively stable patterns over time despite local fluctuations. This result is consistent with findings in human social systems, where interaction patterns also exhibit equilibrium-like properties under bounded adaptability \cite{gonzalez-casado_evidence_2025}. A key limitation of our approach is the simplification introduced by discretizing interaction strengths into just two categories: positive and negative. This coarse-graining may have masked finer-scale dynamics and might have contributed to the apparent equilibrium we observe. We note that employing more refined discretization schemes could help clarify the robustness of these findings.

\subsection{Specific Interactions and the Stress Gradient Hypothesis}

Regarding the dependence of the cooperation-competition ratio and the size of the feasibility domain, the results displayed in Fig.~\ref{fig:ratio-and-feasibility-domain}a highlight how the interplay between environmental change and network structure is consistent with patterns predicted by the Stress Gradient Hypothesis. It suggests that facilitative interactions become more dominant in harsh environments, while competitive interactions prevail under more favourable conditions. This is supported by the fact that, as the proportion of cooperative interactions increases, the coexistence opportunities increase too.

Moreover, this pattern is evident in the \textit{CARACOLES} dataset, as shown by the decreasing trend in the size of the feasibility domain in Fig.~\ref{fig:ratio-and-feasibility-domain}b, which represents a critical state for species coexistence. In contrast, in the \textit{BEEFUN} dataset the feasibility domain stays roughly constant, implying that the wild bee community employs buffering mechanisms: although interactions rearrange, overall coexistence potential is maintained, even if the underlying interactions that produce a feasibility domain differ at each time.

\subsection{The Influence of Environmental Changes on System Stability}


We explored the influence of environmental change on system stability by combining two complementary approaches: first, the analysis of feasible species abundances under hypothetical stationary environmental conditions; and second, the examination of the geometrical properties of the feasibility domain, which reflect the system’s potential to maintain coexistence under demographic perturbations.

In datasets with low temporal resolution --- such as \textit{DIG\_13}, \textit{DIG\_50}, and \textit{LPI\_2858} --- the relationship between environment and species abundance follows a predictable pattern: harsher environmental conditions (e.g., drought or increased temperature) result in smaller feasible abundances, while milder conditions allow for greater population sizes. This inverse relationship is visible in Fig.~\ref{fig:leo-NxP}b. However, from a geometric perspective, both the size of the feasibility domain and the asymmetry index remain constant across environmental conditions. \textit{LPI\_2858} stands out for having the lowest asymmetry index value. This indicates a highly uneven distribution of species extinction probabilities --- some species are much more vulnerable than others. This observation aligns with the patterns found in the feasible abundances, where certain species consistently show much lower expected equilibrium populations. Additional comparisons with \textit{DIG\_13} and \textit{DIG\_50}, provided in the supplementary material, reinforce this conclusion.

For the datasets fitted by means of the full effective adjacency matrix (Eq.~\ref{eq:A_t_model_1}), as seen in Fig.~\ref{fig:asymmetry-index}a, there is no evident simple relation between the feasible species abundances and environmental stress.
Interestingly, we may extract some meaningful information about this \textit{a priori} uncorrelated scenario by means of the asymmetry index. As seen by comparing Fig \ref{fig:ratio-and-feasibility-domain}a) and Fig \ref{fig:leo-NxP}a) for the {\textit{BEEFUN}} dataset, the lower asymmetry is associated to the scenarios in which species abundances are similar, while the higher asymmetry ($J'$ farthest from one) is associated to scenarios in which one species is much more abundant than the others, having a smaller extinction probability. At the level of the size of the feasibility domain, in Fig.~\ref{fig:ratio-and-feasibility-domain}b, we see a tendency for $\Omega$ to decrease as environmental stress grows, except for the last data point ({\textit{CARACOLES}}). This could be explained by the fact that, as mentioned in the methodology Sec.~\ref{subsec:feasibility_domain}, the geometrical measures on stability are conditioned to the hypothesis that population densities thermalize with respect to a given environmental factor. From a phenomenological point of view, this happens when the environmental factors are either very strong or long-lasting. This fact may also help explain the negative density population in the feasible abundancies in Fig.~\ref{fig:leo-NxP} (see Section~\ref{sec:methods} for the methodological discussion).

The asymmetry index proved to be a valuable descriptor of community structure and vulnerability. In datasets modelled via Eq.~\ref{eq:A_t_model_1}, changes in environmental conditions triggered shifts in this index, potentially acting as a precursor to species imbalance or increasing risk of exclusion. By contrast, in the datasets modelled via Eq.~\ref{eq:A_t_model_2}, the asymmetry index remained constant, again highlighting the limitations of static interaction assumptions when compared to time-aware methodologies to represent dynamic ecological realities.

\section{Conclusions}

In this study, we applied a dynamic modelling framework to analyse the temporal evolution of five ecological datasets spanning different species and environments, focusing on the interplay between pairwise interactions and responses to environmental variation. By examining multiple network-level metrics such as rewiring or changes in the sign of the interactions, our analyses provide new insights into the nature of species interactions within each ecological system. This highlights the need for further studies considering temporal changes in the properties of system parameters. 

Lastly, despite the use of relatively simple models in some cases, we were still able to extract meaningful information about the structure and stability of these networks by measuring properties such as the change of size and shape of the feasibility domains in which all species can coexist. These findings suggest that future research incorporating more detailed datasets or refined modelling approaches could further illuminate the mechanisms shaping ecological network structure and their responses to environmental change. 



\subsection*{Acknowledgements}

This work is the outcome of the Complexity72h workshop, held at the Universidad Carlos III de Madrid in Leganés, Spain, from June 23 to 27, 2025. https://www.complexity72h.com. VC-S acknowledges support from the Spanish Ministry of Science and Innovation for the project PID2021-127607OB-I00, and thanks Sergio Picò for providing the seabirds and lizards datasets, Ignasi Bartomeus for providing the wild bee dataset, and Oscar Godoy for providing the annual plant dataset. VC-S especially thanks Francisco P. Molina for collecting and identifying the pollinators of the wild bee dataset, and Lisa Buche and María Hurtado de Mendoza for collecting field data of the annual plant dataset. VHR acknowledges the support of Complexity72h and Coordena\c{c}\~ao de Aperfei\c{c}oamento de Pessoal de N\'ivel Superior (CAPES -- Grant 88887.970397/2024-00). JG-C acknowledges the funding received by Spanish Ministerio de Ciencia, Innovación y Universidades (MICIU/AEI/10.13039/501100011033) through the María de Maeztu project CEX2021-001164-M. AC acknowledges funding by the Maria de Maeztu Programme (MDM-2017-0711) and the AEI (MICIU/AEI/10.13039/501100011033) under the FPI programme.  LSF acknowledges a fellowship from CNPq/Brazil. EE acknowledges the FOCUS TIME CEA. JMH acknowledges support from the Spanish Ministry of Science and Innovation for the projects PID2022-142185NB-C21 and PID2022-142185NB-C22.

\subsection*{Data availability}
The datasets used for this analysis are publicly available in a GitHub repository \footnote{
    \url{https://github.com/violetavivi/complexity72h-2025}.
}.

\clearpage
\bibliographystyle{IEEEtran}
\bibliography{references} 

\end{document}